\documentclass[10pt]{elsarticle}
\journal{ }
\usepackage{amsmath}
\usepackage{indentfirst}
\usepackage{amsfonts,amssymb}
\usepackage{mathrsfs}

\usepackage{amsthm}
\usepackage{lineno,hyperref}
\modulolinenumbers[5]
\newtheorem{Definition}{Definition}[section]
\newtheorem{Theorem}{Theorem}[section]

\numberwithin{equation}{section}
\newtheorem{Remark}{Remark}[section]

\begin{document}
	
	\begin{frontmatter}
\title{ Regulator-based risk statistics with scenario analysis}

\author{Xiaochuan Deng\corref{}}
\address{School of Economics and Management, Wuhan University, 	Wuhan 430072,  China}

\ead{dengxiaochuan@whu.edu.cn}

\author{Fei Sun\corref{mycorrespondingauthor}}
\cortext[mycorrespondingauthor]{Corresponding author}
\address{School of Mathematics and Computational Science, Wuyi University, Jiangmen 529020, China}
\ead{fsun.sci@outlook.com}

\begin{abstract} As regulators pay more attentions to losses rather than gains, we are able to derive a new class of risk statistics, named regulator-based risk statistics with scenario analysis in this paper. This new  class of risk  statistics can be considered as a kind of  risk extension of risk  statistics introduced by Kou et al. \cite{11},  and also data-based versions of loss-based risk measures introduced by  Cont et al. \cite{5} and Sun et al. \cite{12}. 
\end{abstract}


\end{frontmatter}

\section{Introduction}
\label{sec:1}

Research on risk is a hot topic in both quantitative and theoretical research, and risk models have attracted considerable attention. 
The quantitative calculation of risk involves
two problems: choosing an appropriate risk model and allocating the risk to individual institutions. This has led to further research on risk statistics.

In their seminal paper, Artzner et al. \cite{3}\cite{4} first introduced the class of coherent risk measures, by proposing four basic properties to be satisfied by every sound financial risk measure. Further, F\"{o}llmer and Schied \cite{10} and, independently, Frittelli and Rosazza Gianin \cite{9} introduced the broader class, named convex risk measure, by dropping one of the coherency axioms. Later, Hlatshwayo et al. \cite{15} studied the risk measures with bank failure,  Li \cite{16} studied the systemic risk.

However, as pointed out by Cont et al. \cite{5}, these axioms fail to take into account some
key features encountered in the practice of risk management. In fact, sometimes,
when measuring the risk, it is only relevant to consider the losses, not the gains. For this reason, we are able to derive the risk based on losses, not gains.

Next, from the statistical point of view by Kou et al. \cite{11}, the behavior of a random variable can be characterized by its samples. At the same time, one can also incorporate scenario analysis into this framework. Therefore, a natural question is how about the discuss of regulator-based risk with scenario analysis.

In the present paper, we are able to derive convex and coherent regulator-based risk statistics with scenario analysis, and dual representation results for them. Finally, the relationship between  regulator-based risk statistics and the convex risk statistics introsuced by Tian and Suo \cite{14} also be given  to illustrate the regulator-based risk statistics.

It is worth  mentioning  that the issue of risk measures with scenario analysis have already been studied by Delbaen \cite{7}. It have also been extensively studied in the last decade.
For example, see Kou et al. \cite{11},  Ahmed et al. \cite{1},  Assa and Morales \cite{2}, Tian and Jiang \cite{13}, Tian and Suo \cite{14}, and the references therein.  From this point of view, the present paper can also be considered as a kind of  risk extension of risk statistics.

The remainder of this paper is organized as follows. In Sect.~\ref{sec:2}, we briefly introduce some preliminaries. The main results of regulator-based risk statistics be stated in Sect.~\ref{sec:3}, and their proofs be postponed to Sect.~\ref{sec:4}. Finally, in Sect.~\ref{sec:5}, we are able to derive the relationship between regulator-based risk statistics and the convex risk statistics introsuced by Tian and Suo \cite{14},.

\section{Preliminaries}
\label{sec:2}
In this section, we briefly introduce the preliminaries that are used throughout this paper. Let $N \geq 1$ be a fixed positive integer. Denote $\mathscr{X}$ by a set of random losses, and $\mathscr{X}^N$ by the product space
$\mathscr{X}_1 \times \cdots \times\mathscr{X}_N$, where $\mathscr{X}_i=\mathscr{X}$ for $1 \leq i \leq N$. Any element of $\mathscr{X}^N$ is said to be a portfolio of random losses. In practice,
the behavior of the $N$-dimensional  random vector $ {\bf M} = (X_1, \cdots, X_N)$ under different scenarios is represented by
different sets of data observed or generated under those scenarios because specifying accurate models for ${\bf M}$ is usually very difficult.
Some detailed notations can be found in Kou et al. \cite{11}.
Here, we suppose that there always exist $m$ scenarios.
Specifically,   suppose that the behavior of ${\bf M}$
is represented by a collection of data $M=(X_1, \cdots, X_N) \in \mathbf{R}^{N} $ which can be a data set based on historical observations, hypothetical samples simulated according to a model, or a mixture of observations and simulated samples.\\

For any $ M_{1}=(X^1_1, \cdots, X^1_N)$, $  M_{2}=(X^2_1, \cdots, X^2_N)\in \mathbb{R}^{N} $, $ M_{1}\leq M_{2}  $ means $ X^1_i\leq X^2_i $ for any $ i= 1, 2, \cdots, N $.
And for any $M=(X_1, \cdots, X_N) \in \mathbf{R}^{N} $, let $M \wedge 0 := \big(\min \{X_{1}, 0\}, \cdots, \min \{X_{N}, 0\}\big)$. Given  $ a\in  \mathbf{R} $, denote $ a1:= \big(a, \cdots, a\big) $.

\section{Regulator-based risk statistics}
\label{sec:3}
In this section, we state the main result of regulator-based risk statistics with scenario analysis. Firstly, we derive the properties related to regulator-based risk statistics.

\begin{Definition}
		A function $\rho: \mathbb{R}^N\rightarrow [0,+\infty)$ is said to be a convex regulator-based risk statistic if it satisfies the following properties, \\
		(A.1)Normalization: for cash losses: for any $a \geq0$, $\rho(-(a1))=a;$\\
		(A.2)Monotonicity: for any $ M_1, M_2 \in \mathbb{R}^N$, $M_1\leq M_2$ implies $\rho(M_1)\geq \rho(M_2);$\\
		(A.3)Loss-dependence : for any $M \in \mathbb{R}^N$, $\rho(M)=\rho(M\wedge0)$.\\
		(A.4)Convexity:  for any $ M_1, M_2 \in \mathbb{R}^N$ and $0 <\lambda<1$,
		$\rho(\lambda M_1+ (1-\lambda) M_2)\leq \lambda \rho(M_1) +(1-\lambda)\rho(M_2).$\\
Moveover, a convex regulator-based risk statistic $\rho$ is said to be a coherent regulator-based risk statistic if it still satisfies\\
(A.5)Positive homogeneity: for any $\alpha \geq 0$ and $M\in \mathbb{R}^N$, $\rho(\alpha M)=\alpha\rho(M).$
\end{Definition}

Next, we derive the dual representations of regulator-based risk statistics and the proofs were leaved in next section.

\begin{Theorem}\label{T31}
$\rho:\mathbb{R}^N \rightarrow [0, +\infty)$ is a convex regulator-based risk statistic in the case of that there exists a convex function $\alpha:\mathbb{R}^N\rightarrow [0,+\infty]$,
which is satisfied
\begin{equation}\label{33}
\min\limits_{Q \in \mathbb{R}^N, \min{Q_i}\geq 1-\epsilon} \alpha(Q) =0 \quad \text{ for any } \epsilon \in (0,1)
\end{equation}
such that
\begin{equation}\label{31}
\rho(M)=\max\limits_{Q\in \mathbb{R}^N}\{-\sum\limits_{i=1}^N{Q_i}(X_i\wedge0)-\alpha(Q)\}.
\end{equation}
The function $\alpha$ for which (\ref{31}) holds can be choose as
$\alpha_{\min}(Q):= \sup\limits_{M \in \mathbb{R}^N}\{-\sum\limits_{i=1}^N {Q_i}(X_i)- \rho(M)\}$
for any $Q \in \mathbb{R}^N$. Moreover, $\alpha_{\min}$ is the minimal penalty function in the sense that for any penalty function $\alpha$ representing $\rho$ satisfies $\alpha(Q_1, \cdots, Q_N) \geq \alpha_{\min}(Q_1, \cdots, Q_N)$ for all $(Q_1, \cdots, Q_N) \in \mathbb{R}^N$.
\end{Theorem}

\begin{Theorem}\label{T32}
$\rho:\mathbb{R}^N \rightarrow [0, +\infty)$ is a coherent regulator-based risk statistic in the case of that for any $M\in \mathbb{R}^N$, 
\begin{equation}\label{32}
\rho(M)=\max\limits_{Q\in \mathbb{R}^N}\{-\sum\limits_{i=1}^N{Q_i}(X_i\wedge0)\}.
\end{equation}
\end{Theorem}
\begin{Remark}
 The dual representation result in Theorem~\ref{T31} depends only on the negative part of $M$ due to the loss-dependence property (A.3).
 In Theorem~\ref{T32}, let $N=1$, then representation result is reduced to the one-dimensional case which coincides with the representation results of Cont et al. \cite{5}. 
\end{Remark}

\section{Proofs of main results}
\label{sec:4}
In this section, we are able to derive the proof of main results in Sect.~\ref{sec:3}.\\

\noindent \textbf{Proof of Theorem~\ref{T31}.}
 Let $f(X)=\rho(-X)$, then $f$ is an increasing convex function. According to [Cheridto and Li \cite{6}, Th4.2], we have
$$f(M)=\max\limits_{M^* \in \mathbb{R}^N}\{M^*(M)-f^*(M^*)\}$$
where
$$f^*(M^*)=\sup\limits_{M \in \mathbb{R}^N}\{M^*(-M)-\rho(M)\}.$$
Hence
$$\rho(M)=f(-M)=\max\limits_{M^* \in \mathbb{R}^N}\{M^*(-M)-f^*(M^*)\}.$$
Hence
$$\rho(M)=\max\limits_{Q \in \mathbb{R}^N}\{-\sum\limits_{i=1}^N{Q_i}(X_i)-f^*(Q)\},$$
where
$$ f^*(Q)=\sup\limits_{Q \in\mathbb{R}^N} \{-\sum\limits_{i=1}^N{Q_i}(X_i)- \rho(M)\}.$$
Define  $\alpha_{\min}:\mathbb{R}^N\rightarrow [0,+\infty]$ by
$$\alpha_{\min}(Q):= \sup\limits_{Q \in \mathbb{R}^N} \{-\sum\limits_{i=1}^N{Q_i}(X_i)- \rho(M)\},$$
and using  loss-dependence property of $\rho$, we have
$$\rho(M)=\max\limits_{Q \in \mathbb{R}^N}\{-\sum\limits_{i=1}^N{Q_i}(X_i\wedge 0)-\alpha_{\min}(Q)\}.$$
Now, let $\alpha$ be any penalty function for $\rho$. Then, for any $(Q_1, \cdots, Q_N) \in \mathbb{R}^N$ and $ M =(X_1, \cdots, X_N),$
$$\rho(M) \geq -\sum\limits_{i=1}^N {Q_i}(X_i) - \alpha(Q_1, \cdots, Q_N).$$
Hence,
$$\alpha(Q_1, \cdots, Q_N) \geq -\sum\limits_{i=1}^N {Q_i}(X_i) - \rho(M).$$
Taking supremum over $\mathbb{R}^N$ for $M=(X_1, \cdots, X_N)$ in give rise to
\begin{align*}\alpha(Q)&\geq \sup\limits_{(X_1, \cdots, X_N)\in \mathbb{R}^N}\{-\sum\limits_{i=1}^N {Q_i}(X_i) - \rho(M)\}\\
&=\alpha_{min}(Q)
\end{align*}
Next, we check that $\rho$ represented in (\ref{31}) is a convex regulator-based risk statistic.
Obviously, $\rho$ is a
convex function and satisfies (A.3). Hence, we need only to show that $\rho$ satisfies (A.1) and (A.2). To this end, for any $a\geq 0$ and $ 1< \epsilon < 1 $,
\begin{align*}
a &=\rho\left(-a1\right)\\
&=\max\limits_{Q \in \mathbb{R}^N}\{a\sum\limits_{i=1}^N{Q_i}-\alpha_{\min}(Q)\}\\
&\leq\max\Big\{\max\limits_{Q \in \mathbb{R}^N, \max\limits_{1\leq i\leq N} Q_i< 1-\epsilon}\{a\sum\limits_{i=1}^N{Q_i}-\alpha_{\min}(Q)\},  \max\limits_{Q \in \mathbb{R}^N, \min\limits_{1\leq i\leq N} Q_i\geq 1-\epsilon}\{a\sum\limits_{i=1}^N{Q_i}-\alpha_{\min}(Q)\}, \\
&\quad \quad  \max\limits_{Q \in \mathbb{R}^N, \min\limits_{1\leq i\leq N} Q_i\leq 1-\epsilon, 
	\max\limits_{1\leq i\leq N} Q_i \geq 1-\epsilon}\{a\sum\limits_{i=1}^N{Q_i}-\alpha_{\min}(Q)\}\Big\}\\
&\leq\max\Big\{Na(1-\epsilon),  \max\limits_{Q \in \mathbb{R}^N, \min\limits_{1\leq i\leq N} Q_i\geq 1-\epsilon}\{a\sum\limits_{i=1}^N{Q_i}-\alpha_{\min}(Q)\}, \\
&\quad \quad  \max\limits_{Q \in \mathbb{R}^N, \min\limits_{1\leq i\leq N} Q_i\leq 1-\epsilon,
	\max\limits_{1\leq i\leq N} Q_i \geq 1-\epsilon}\{a\sum\limits_{i=1}^N{Q_i}-\alpha_{\min}(Q)\}\Big\}\\
&\leq\max\Big\{Na(1-\epsilon), a - \min\limits_{Q \in \mathbb{R}^N,
	\min\limits_{1 \leq i \leq N}Q_i\geq 1-\epsilon} \alpha_{\min}(Q), \\
& \quad \quad card(i \in \{i: Q_i(1)< 1-\epsilon\}) a (1-\epsilon)+ card(i \in \{i: Q_i(1)\geq 1-\epsilon\})a -\\
& \quad \quad \min\limits_{Q \in \mathbb{R}^N,\min\limits_{1\leq i\leq N} Q_i< 1-\epsilon, \max\limits_{1\leq i\leq N} Q_i > 1-\epsilon} \alpha_{\min}(Q)\Big\},
\end{align*}
which implies $ \alpha_{\min} $ satisfies (\ref{33}).
Now, let $M_1:=(X^1_1, \cdots, X^1_N),  M_2:=(X^2_1, \cdots, X^2_N)$. Then, the relation $M_1 \leq M_2$ implies  $X^1_i \wedge 0 \leq X^2_i \wedge 0$ for any $1 \leq i \leq N$.
Hence for any $Q:=(Q_1, \cdots, Q_N) \in \mathbb{R}^N$, we have
$$\sum\limits_{i=1}^N {Q_i}(X^1_i \wedge 0)\leq \sum\limits_{i=1}^N {Q_i}(X^2_i \wedge 0),$$
which implies $\rho(M_1) \leq \rho(M_2).$
This completes the proof of Theorem~\ref{T31}.\qed\\

\noindent \textbf{Proof of Theorem~\ref{T32}.}
If $\rho$ is a coherent regulator-based risk statistic, then from the proof of Theorem 3.1 and the positive homogeneity of $\rho$, for any $Q \in \mathbb{R}^N$ and $\lambda > 0$, we have
\begin{align*}\alpha_{\min}(Q)&=\sup\limits_{M \in \mathbb{R}^N} \{-\sum\limits_{i=1}^N{Q_i}(-X_i)-\rho(M)\}\\
   &=\sup\limits_{M \in \mathbb{R}^N} \{-\sum\limits_{i=1}^N{Q_i}(-\lambda X_i)-\rho(\lambda M)\}\\
      &=\lambda \sup\limits_{M \in \mathbb{R}^N} \{-\sum\limits_{i=1}^N {Q_i}(-X_i)-\rho(M)\}\\
        &=\lambda \alpha_{\min} (Q)
          \end{align*}
Hence, $\alpha_{\min}$ can take only the values $0$ and $+\infty$. This completes the proof of Theorem~\ref{T32}.\qed\\

\section{Regulator-based version of convex risk statistics}
\label{sec:5}
In this section, we derive a new version of regulator-based risk statistics. It is worth noting that this version can be related to convex risk statistics introsuced by Tian and Suo \cite{14},.\\

For any convex risk statistic $\bar{\rho}$ on  $\mathbb{R}^N$ defined in Tian and Suo \cite{14},, we can define a new risk statistic $\rho$
by $\rho(M):=\bar{\rho}(M \wedge 0)$ for any $M \in \mathbb{R}^N$. Obviously, $\rho$ is a convex regulator-based risk statistic defined in Sect.~\ref{sec:3}. We call $\rho$ the regulator-based version of $\bar{\rho}$.\\

We can prove that a convex regulator-based risk statistic $\rho$ is the regulator-based version of some convex risk statistic in the case of that it satisfies

Cash-loss additivity: for any $M \in \mathbb{R}^N$ and $ a\in \mathbb{R} $ where $M \leq 0$, $a \geq 0$, \[\rho\left(M-a1\right)=\rho(M)+a.\]

On the one hand, if $\rho(M)=\bar{\rho}(M\wedge 0)$ for certain convex risk statistic $\bar{\rho}$ on $\mathbb{R}^N$, then for any $M \in \mathbb{R}^N$,
$M \leq 0$ and $a \geq 0$,
$$\rho\left(M-a1\right)=\bar{\rho}\left(M-a1\right)=\bar{\rho}(M)+a=\rho(M)+a.$$
where the second equality is due to the cash-additivity property of $\bar{\rho}$.\\

Let us now suppose that a convex regulator-based risk statistic $\rho$ satisfies the cash-loss additivity property. Define
$$\bar{\rho}(M)=\rho\Big(M- a_{M}1\Big)-a_{M}$$
for any $M:=(X_1, \cdots,X_N) \in \mathbb{R}^N$
where $a_{M}$ is any upper-bound of each $X_i$. Using the cash-loss additivity property for $\rho$, we know that $\bar{\rho}$ is well-defined. Next,  we need to claim that $\bar{\rho}$ is a convex risk statistic  where $\rho(M)=\bar{\rho}(M \wedge 0)$. To this end, for any $M :=(X_1, \cdots, X_N)\in \mathbb{R}^N$ and $a \in \mathbb{R}$,
\begin{align*}
\bar{\rho}\left(M-a1\right)&=\rho\left(M-a1-(a_{M}1-a1)\right)-(a_{M}-a)\\
&=\rho\left(M-a_{M}1\right)-a_{M}+a\\
&=\bar{\rho}(M)+a.
\end{align*}
Next, let $M_1:=(X_1^1,\cdots,X_N^1), M_2:=(X_1^2,\cdots,X_N^2) \in \mathbb{R}^N$ where $M_1 \leq M_2$. Taking $a_{M_1}$, $a_{M_2}$ to be the upper-bound of each $X^1_i$ and $X^2_i$.
Then, 
\begin{align*}
\bar{\rho}(M_1)&=\rho(M_1-a_{M_2}1)-a_{M_2}\\
&\geq \rho(M_2-a_{M_2}1)-a_{M_2}\\
&=\bar{\rho}(M_2),
\end{align*}
which yields $ \bar{\rho} $ is monotonous.
Finally, for any $ M_1, M_2 \in \mathbb{R}^N$ and $ 0\leq t \leq 1 $,
 \begin{align*}\bar{\rho}\left(t M_1+(1-t)M_2\right)&=\rho\Big(tM_1+(1-t)M_2-ta_{M_1}1-(1-t)a_{M_2}1\Big)- ta_{M_1}- (1-t)a_{M_2}\\
   &=\rho\Big(t (M_1-a_{M_1}1)+(1-t)(M_2- a_{M_2}1)\Big)- ta_{M_1}- (1-t)a_{M_2}\\
    &\leq t\rho\Big( M_1-a_{M_1}1\Big)+(1-t)\rho\Big(M_2- a_{M_2}1\Big)- ta_{M_1}- (1-t)a_{M_2}\\
    &=t\bar{\rho}(M_1)+(1-t)\bar{\rho}(M_2),
     \end{align*}
which implies $ \bar{\rho} $ is convex.

\end{document}